\documentclass{article}
\usepackage {graphicx}
\begin{document}
\baselineskip .75cm 
\begin{titlepage}
\title{\bf Thermodynamics of (2+1) flavor QGP in quasiparticle model}       
\author{Vishnu M. Bannur  \\
{\it Department of Physics}, \\  
{\it University of Calicut, Kerala-673 635, India.} }   
\maketitle
\begin{abstract}

Using our recently developed one parameter quasiparticle model, we analyze more recent (refined) results of (2+1) flavor QGP in lattice simulation of QCD by various groups \cite{f.1,k.1,u.1,d.1}. We got a  remarkable good fit to lattice thermodynamics of Ref. \cite{f.1} and reasonable good fit to Ref. \cite{k.1} by adjusting single parameter of the model which may be related QCD scale parameter. Further we extend our model for above system with zero chemical potential to non-zero chemical potential and predict quark density without any new parameters which may be compared with future lattice data.   
\end{abstract}
\vspace{1cm}
                                                                                
\end{titlepage}
\section{Introduction :}
Recently, there are various attempts to study lattice quantum chromodynamics (QCD) with quarks by simulations at finite temperature and chemical potential which describes a system with quarks and gluons in plasma state, called quark gluon plasma (QGP), at thermodynamic equilibrium and various thermodynamic quantities are derived. Thermodynamics of QGP is important to describe the evolution of fire-ball formed in relativistic heavy ion collisions, astrophysical objects and early universe so on. It is a challenging job to minimize lattice artefacts and obtain a reliable results. Eventhough results of various groups \cite{f.1,k.1,u.1,d.1} predict nonideal effects qualitatively, there were differences in quantitative results because of lattice artefacts. Therefore, it may be useful to analyze lattice gauge theory (LGT) results of various groups with theoretical models. 

Quasiparticle model is one of the simplest model, widely studied in literature \cite{s.1,p.1,p.2}. There are varieties of quasiparticle models, but all of them with few phenomenological parameters. We have also developed a quasiparticle model with only one parameter \cite{ba.1}. We use this model to study, in a unified manner, the LGT results of various groups and a reasonable good fits were obtained to all results by adjusting the single parameter which may be related QCD scale parameter. Difference  between various lattice results reflects as different values for the parameter of the model.  

Next section we discuss the concept of statistical mechanics of quasiparticle model. In section III quasiparticle model for (2+1) flavor QGP is discussed. Results and conclusions follows in section IV. 

\section{Quasiparticle Model:} 

It is a phenomenological model to describe the statistical mechanics and thermodynamics of a non-ideal system. It is assumed that the effects of interaction leads to thermal excitation of quasiparticles with thermal mass which is related to interaction strength or  coupling constant and temperature. We assume that whole thermal energy is used to excite this quasiparticles and it is an ideal system of quasiparticles. It means that system with real particles with mutual interaction at thermodynamic equilibrium is approximated by ideal system with quasiparticles at thermodynamic equilibrium. However, all results of standard statistical mechanics of ideal system may not be applicable here as such, because of the temperature dependent thermal masses of quasiparticles. 

Suppose we consider such a system in canonical ensemble, then we know that the system is in thermal contact with a reservoir and attains the temperature of reservoir at equilibrium. Partition function is \cite{pa.1} 
\begin{equation}
Q_N = \sum_r e^{- \beta E_r} ,  
\end{equation}
and average energy, by definition, 
\begin{equation}
U = <E_r> = - \frac{\partial \,ln Q_N}{\partial \beta} ,\label{eq:u}
\end{equation}
where $\beta \equiv 1/T$ and $E_r$ is the $r^{th}$ energy state of the system. In quasiparticle model $E_r$ also depends on $T$ through the mass of quasiparticles and Eq.(\ref{eq:u}) is just an identity, namely, explicit derivative of partition function with respect to $\beta$. In terms of T, above equation becomes 
\begin{equation}
U = T^2 \frac{d \,ln Q_N}{d T} - T^2 \frac{1}{Q_N} \sum_r e^{- \beta E_r} \frac{\partial E_r}{\partial T} ,
\end{equation}
which may be written as 
\begin{equation}
\frac{U}{T^2} = \frac{d }{d T} \left( ln Q_N - \int_{T_0}^T d\tau \tau^2 \frac{\partial m(\tau)}{\partial \tau} <\frac{\partial E_r}{\partial m}> \right) \, \label{eq:ut}
\end{equation}
Other thermodynamic quantities are obtained from the Helmholtz free energy,  
\begin{equation}
A = U - T S = U + T \frac{\partial A}{\partial T} ,  
\end{equation}
which may be written as 
\begin{equation}
\frac{U}{T^2} = \frac{d }{d T} (\frac{A}{T}) .
\end{equation}
Comparing with Eq.(\ref{eq:ut}), we get,
\begin{equation}
A = - T ln Q_N + \int_{T_0}^T d\tau \tau^2 \frac{\partial m(\tau)}{\partial \tau} <\frac{\partial E_r}{\partial m}> . \label{eq:a} 
\end{equation}
This is known as the bridging relationship between statistical mechanics (right hand side) and thermodynamics (left hand side). Since the Helmholtz free energy is a thermodynamic quantity, it is a function of temperature (T), Volume (V) and number of particles (N) and supposed to be unaware of the temperature dependence of quasiparticle mass. Quasiparticle mass enters only through statistical mechanics, right hand side of Eq.(\ref{eq:a}). All other thermodynamic quantities may be obtained from $A(T,V,N)$ using thermodynamic relations. For example, pressure is given by 
\begin{equation}
p = - \frac{\partial A}{\partial V} = p_{id} + f(T), 
\end{equation} 
where $p_{id}$ is the expression for pressure of ideal system. Thus we get an extra  temperature dependent term $f(T)$ to pressure compared to ideal gas. Many earlier authors \cite{p.1} and even today \cite{v.1}, overlook this term and neglect $f(T)$ and lead to thermodynamic inconsistency and so on. Of course, there are attempts by \cite{fa.1} to unify all different models in a thermodynamically self-consistent models. We still believe that our approach is a natural extention of normal statistical mechanics to temperature dependent quasiparticle system, mainly because left hand side  of Eq. (\ref{eq:ut}, \ref{eq:a}), being a thermodynamic function, should not explicitly depend on $m(T)$, but over all dependence on temperature. Whereas statistical mechanics, right hand side depends on $m(T)$. Therefore, in our analysis we first evaluate U or $\varepsilon \equiv \frac{U}{V}$, energy density and derive pressure and other thermodynamic quantities using thermodynamic relations.   

\section{(2+1) flavor QGP:}

This system is assumed to be consists of  gluons, up and down quarks, and strange quarks where all of them treated as non-interacting quasiparticles with respective temperature dependent thermal masses in addition to physical masses. Following the standard text book statistical mechanics \cite{pa.1}, we calculate $\varepsilon(T)$ first and then $p(T)$ is obtained from thermodynamic relations. Statistical mechanics and hence $m(T)$ enter only in $\varepsilon(T)$. Treating as grand canonical ensemble with zero chemical potential we have,
\begin{equation}
\frac{\varepsilon_g}{T^4} = \frac{d_g}{2 \pi ^2} \int_0^{\infty} dx\frac{ x^2 \sqrt{x^2 + \bar{m}_g^2}}{e^{\sqrt{x^2 + \bar{m}_g^2}} -1} , 
\end{equation}
for gluons and 
\begin{equation}
\frac{\varepsilon_q}{T^4} = \frac{d_q}{2 \pi ^2} \int_0^{\infty} dx\frac{ x^2 \sqrt{x^2 + \bar{m}_q^2}}{e^{\sqrt{x^2 + \bar{m}_q^2}} +1} , 
\end{equation}
for quarks. $d_g = 16$ and $d_q = 12$ are the internal degrees of freedom of gluons and quarks. $\bar{m}_g \equiv m_g/T$ and $\bar{m}_q \equiv m_q/T$, where $m_g$ and $m_q$ are the total masses of gluons and quarks, respectively, including thermal masses. For (2+1) flavor QGP, total energy density is 
\begin{equation}
\varepsilon = \varepsilon_g + 2 \varepsilon_u + \varepsilon_s  , \label{eq:qgp}
\end{equation}  
where u and s refers light (up and down) quarks and strange quark, respectively. 
Gluon mass may be taken as \cite{p.2,ba.1},
\begin{equation}
m_g^2 = \frac{3}{2} \omega_p^2 ,
\end{equation} 
and quark mass
\begin{equation}
m_q^2 = (m_{q0}+ m_f)^2 + m_f^2 
\end{equation} 
Plasma frequencies $\omega_p$ and $m_f$ arise because of the collective behavior of plasma due to mutual interactions among particles. In the absence of interaction, whole thermal energy may be used to excite real particles from the vacuum. From the perturbative QCD calculations we have 
\begin{equation}
\omega_p^2 = \frac{g^2 T^2}{18} (2 N_c) + \frac{g^2 T^2}{18} n_f  . \label{eq:op}
\end{equation}   
where $N_c =3$ for QCD and $n_f$ is the number of flavors. $g$ is the coupling constant.   
Similarly, 
\begin{equation}
m_f^2 = \frac{(N_c^2 - 1)}{2 N_c} \frac{g^2 T^2}{8} . \label{eq:oq}
\end{equation}
Above expressions for plasma frequencies are valid at high temperature. To include low temperature as well, we model them as, 
\begin{equation}
\omega_p^2 = a_g^2 \,g^2 \,\frac{n_g}{T}+ \sum_q a_q^2 \,g^2 \,\frac{n_q}{T} , \label{eq:wp}
\end{equation}
and 
\begin{equation}
m_f^2 = b_q^2 \,g^2 \,\frac{n_q}{T}  \label{eq:mf} 
\end{equation}
such that at high temperature they goes to QCD perturbative results Eqs.(\ref{eq:op}, \ref{eq:oq}) and the  coefficients $a_g$, $a_q$ and $b_q$ may be determined. $n_g$ and $n_q$ are densities of gluons and quarks respectively and defined as,
\begin{equation}
\frac{n_g}{T^3} = \frac{d_g}{2 \pi ^2} \int_0^{\infty} dx\frac{ x^2 }{e^{\sqrt{x^2 + \bar{m}_g^2}} -1} , \label{eq:ng}
\end{equation}
for gluons and 
\begin{equation}
\frac{n_q}{T^3} = \frac{d_q}{2 \pi ^2} \int_0^{\infty} dx\frac{ x^2 }{e^{\sqrt{x^2 + \bar{m}_q^2}} +1} , \label{eq:nq}
\end{equation}
for quarks. 
\begin{figure}[h]
\centering
\includegraphics[height=8cm,width=12cm]{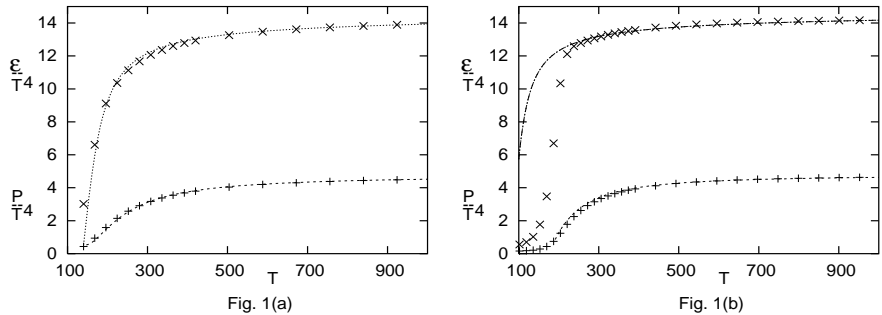}
\caption{ Plot of $\varepsilon/T^4$ and $P/T^4$ as a function of $T$ for (2+1) flavor QGP from our quasiparticle model and compared with LGT results of Ref. \cite{f.1} (symbols), Fig. 1(a), and Ref. \cite{k.1}, Fig. 1(b).} 
\end{figure}
  
Above form for frequencies is motivated by similar work in ultra-relativistic $(e,e^+,\gamma)$  plasma \cite{m.1, ba.2}, where plasma frequencies are proportional to $\frac{n e^2}{T}$. This method works well to explain LGT results of pure gluon, 2 flavor, 3 flavor and earlier (2+1) flavor QGP LGT results \cite{ba.1}. $\frac{n_g}{T^3}$ and $\frac{n_q}{T^3}$ are obtained from it's definition Eqs.(\ref{eq:ng}, \ref{eq:nq})  which depends again on thermal masses and hence becomes integral equation which need to be solved self-consistently and which on substitution in Eq. (\ref{eq:qgp}) we get energy density. Then all other thermodynamics may be obtained from thermodynamic relations. We use 2-loop running coupling constant, 
\begin{equation} \alpha_s (T) \equiv \frac{g^2}{4 \pi} = \frac{6 \pi}{(33-2 n_f) \ln (T/\Lambda_T)}  
\left( 1 - \frac{3 (153 - 19 n_f)}{(33 - 2 n_f)^2} 
\frac{\ln (2 \ln (T/\Lambda_T))}{\ln (T/\Lambda_T)} \right)  \;, \label{eq:as} 
\end{equation}
which appear in thermal masses. Only parameter in the model is the $\Lambda_T$ which we adjust to get the best fit to LGT results. 
\begin{figure}[h]
\centering
\includegraphics[height=8cm,width=12cm]{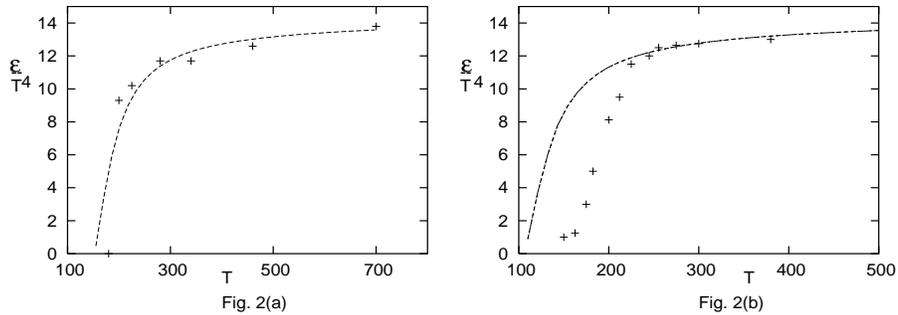}
\caption{ Plot of $\varepsilon/T^4$ as a function of $T$ for (2+1) flavor QGP from our quasiparticle model and compared with LGT results (symbols) of Ref. \cite{u.1}, Fig. 2(a) and Ref. \cite{d.1}, Fig. 2(b).} 
\end{figure}
   
\section{Results and Conclusions:} 

First, we have to solve equations for densities of up quarks, strange quarks and then gluons, separately, which are all integral equations, Eq. (\ref{eq:nq}, \ref{eq:ng}), and then these densities are used to get the thermal masses, Eq. (\ref{eq:wp}, \ref{eq:mf}).  Using these thermal masses energy density, Eq. (\ref{eq:ut}), is calculated. Only parameter in the model is $\Lambda_T$ which we adjust such that we get good fit to LGT results. However, we found that LGT data of \cite{k.1} lies above our results and that of \cite{f.1} lies below our results. Hence, we multiply LGT data with appropriate factor, as done in earlier quasiparticle models models \cite{p.2} which may be accounted for uncertainties in LGT data. In Fig. 1(a), we compare our results with LGT results of Ref. \cite{f.1}.  Physical masses of strange quark is taken as 150 MeV and light quark mass is $\frac{1}{28.15}^{th}$ of strange quark mass. To get good fit we need $\Lambda_T = 135$ MeV and LGT data need to be multiplied by a factor  1.07. Similar comparison is made for LGT data of Ref. \cite{k.1} in Fig. 1(b) and good fit is obtained for $\Lambda_T = 80$ MeV and with LGT data multiplied by a factor .99. Strange quark mass is taken as 150 MeV and light quark mass is $\frac{1}{10}^{th}$ of strange quark mass. We see that two different LGT data lie in two opposite sides of quasiparticle model. The discrepency in two different LGT data was attributed to uncertainties associated with lattice artefacts. For both the LGT results we took the respective parameterized functions presented in references \cite{f.1,k.1} which reproduces LGT data. In Fig. 2, we also compare our model with preliminary results of \cite{u.1} and \cite{d.1}, and fitting parameters are $\Lambda_T = 150$ MeV with  multiplication factor 1.2 and $\Lambda_T = 105$ MeV without any multiplication factor, respectively. Pressure ($P/T^4$) is obtained from energy density using thermodynamic relations
 \begin{equation}
\varepsilon =  T \frac{\partial P}{\partial T} 
- P \,\, ,  \label{eq:td} 
\end{equation}
on integration over temperature as done in Ref. \cite{ba.1}. Since quasiparticle model is valid for plasma state, lower limit of integration is taken to be critical temperature. We see from Fig. 1(a) that quasiparticle model fits LGT data of Ref. \cite{f.1} surprisingly well for both the energy density and pressure. In Fig. 1(b) we have similar  comparison with LGT results of Ref. \cite{k.1} and obtain a reasonably good fit except very near to $T_c$. For other LGT data our results are compared in Fig. 2(a) for Ref. \cite{u.1} and Fig. 2(b) for Ref. \cite{d.1}. In these two cases we don't have any parameterized functions to represent LGT data to fine tune the parameter of our model. 
\begin{figure}[h]
\centering
\includegraphics[height=8cm,width=12cm]{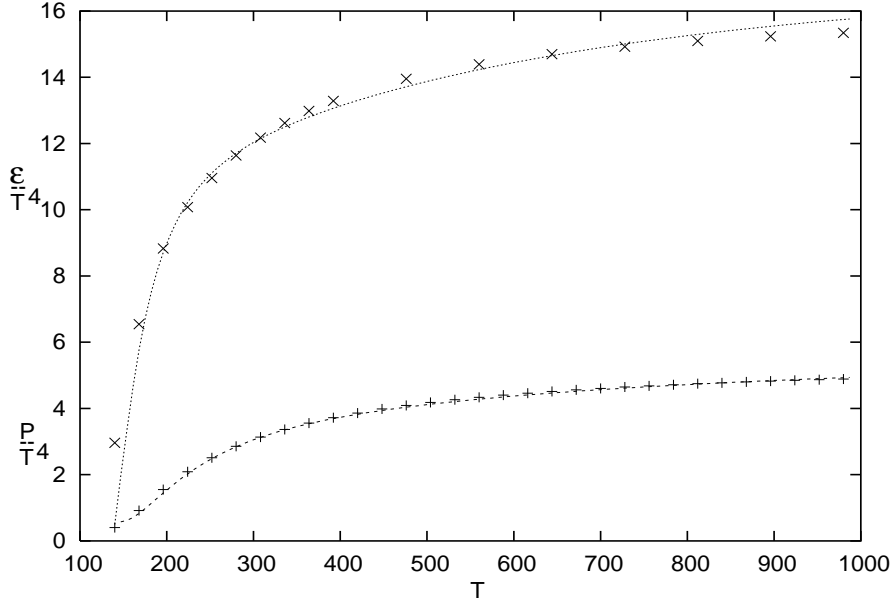}
\caption{ Plot of $\varepsilon/T^4$ and $P/T^4$ as a function of $T$ for (2+1+1) flavor QGP from our quasiparticle model and compared with LGT results of Ref. \cite{f.1} (symbols).} 
\end{figure}

It is interesting to note that our model, extended to (2+1+1) flavor QGP by including charm quarks with physical mass equal to 11.85 times the strange quark mass, also fits the corresponding LGT data of Ref. \cite{f.1} reasonably well as shown in Fig. 3. Here $\Lambda_T = 135$ MeV and multiplication factor is .93.  

So far we concentrated on a system with zero chemical potential and next we extend the model to system with finite chemical potential. Following the literature \cite{p.2,ba.1}, we just need to replace $T/\Lambda_T$ in coupling constant, Eq. (\ref{eq:as}), by 
\begin{equation} 
\frac{T}{\Lambda_T} \,\sqrt{1 +  (1.91/2.91)^2 \,\frac{\mu^2}{T^2} } \,\,,  
\label{eq:lsa} 
\end{equation} 
using the results of Schneider \cite{sc.1,ba.1} and replacing perturbative plasma frequencies by,  
\begin{equation}
m_f^2 = \frac{g^2\,T^2}{6} (1 + \frac{\mu_f^2}{\pi^2 \,T^2} )\;\;, 
\end{equation}
and  
\begin{equation}
\omega_p^2 = \frac{g^2\,T^2}{18}\,((6 + n_f) + \frac{3 \sum_f \mu_f^2}{\pi^2\,T^2})\;\;, 
\end{equation}
where $\mu_f$ is the quark chemical potential of different flavors. Here we consider a system with $\mu_u = \mu_d \equiv \mu$ and all other chemical potentials are zero. With these modifications, but without any new parameters, light quark density, $N_q \equiv (n_l - \bar{n}_l)$, is plotted in Fig. 3(a) and Fig. 3(b) corresponding the fitting parameters of Ref. \cite{f.1} and Ref. \cite{k.1} respectively which may be compared with lattice results in future. There is a wide difference between two results except at very high temperature. $n_l$, $\bar{n}_l$ refers to light (up and down) quark and antiquark densities. 
\begin{figure}[h]
\centering
\includegraphics[height=8cm,width=12cm]{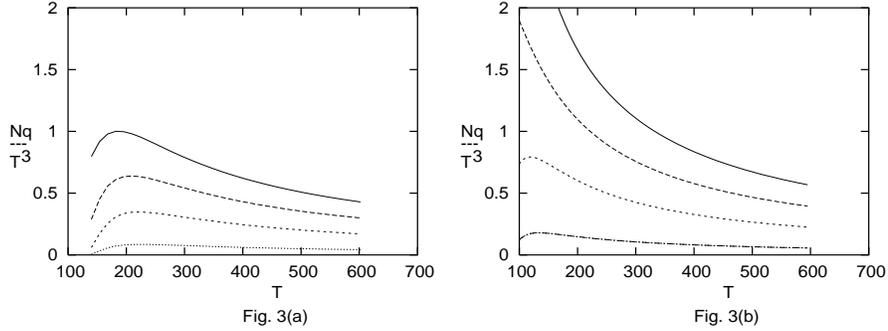}
\caption{ Plot of $N_q/T^3$ as a function of $T$ for (2+1) flavor QGP from our quasiparticle model with the fitting parameters of Ref. \cite{f.1}, Fig. 3(a) and \cite{k.1}, Fig. 3(b) for $\mu/T_c$ equal to 1.0, 0.7, 0.4 and .1 (top to bottom curves respectively).} 
\end{figure}  

In conclusion, we applied our one parameter quasiparticle model to (2+1) flavor QGP system and studied the statistical mechanics and thermodynamics, and compared with lattice data of various groups. It is remarkable that just by adjusting one parameter, QCD scale parameter of 2-loop coupling constant, reproduces lattice results of various groups and it's values lie between 80 to 150 MeV. No need of use of full LGT data of one of the thermodynamic quantity to obtain model function, as done in some quasiparticle models, and then evaluating other thermodynamic quantities and fitting to corresponding LGT data. This corresponds to infinite parameter model and definitely will fit LGT data because of thermodynamic relations. Shortcomings of our model may be the ignorance of the QCD parameter $\Lambda_T$ and the origin of multiplication factors which are different for different lattice groups. Of course, almost all other models used to describe the non-ideal behaviour of QGP have more than one parameter.         

\end{document}